\documentclass[11pt,twocolumn]{article}

\usepackage{graphicx}
\usepackage[top=1in, bottom=1.1in, left=0.8in, right=0.8in]{geometry}
\usepackage[super,sort,compress]{cite}

\title{\textbf{Molecular Simulations of the Ribosome and Associated Translation Factors}}

\author{
    Lars V. Bock
    \and
    Michal H. Kol\'a\v{r}
    \and
    Helmut Grubm\"uller$^*$
}

\date{}

\begin{document}

\twocolumn[

    \begin{@twocolumnfalse}
    
    \maketitle
    
    \centering{
        \begin{minipage}{0.9\textwidth}
            \centering{
                \textit{Department of Theoretical and Computational     Biophysics,\\  Am Fa{\ss}berg 11, G\"ottingen, Germany\\
                $^*$corresponding author: hgrubmu@gwdg.de}
            }
        \end{minipage}
    }
    
    \vspace{2em}
    
    \centering{
        \begin{minipage}{0.9\textwidth}
            The ribosome is a macromolecular complex which is responsible for protein synthesis in all living cells according to their transcribed genetic information. Using X-ray crystallography and, more recently, cryo-electron microscopy (cryo-EM), the structure of the ribosome was resolved at atomic resolution in many functional and conformational states. Molecular dynamics simulations have added information on dynamics and energetics to the available structural information, thereby have bridged the gap to the kinetics obtained from single-molecule and bulk experiments. Here, we review recent computational studies that brought notable insights into ribosomal structure and function.
        \end{minipage}
    }

    \vspace{2em}
    \end{@twocolumnfalse}
]

\section*{Introduction}

\begin{figure*}
\begin{center}
\includegraphics[width=\textwidth]{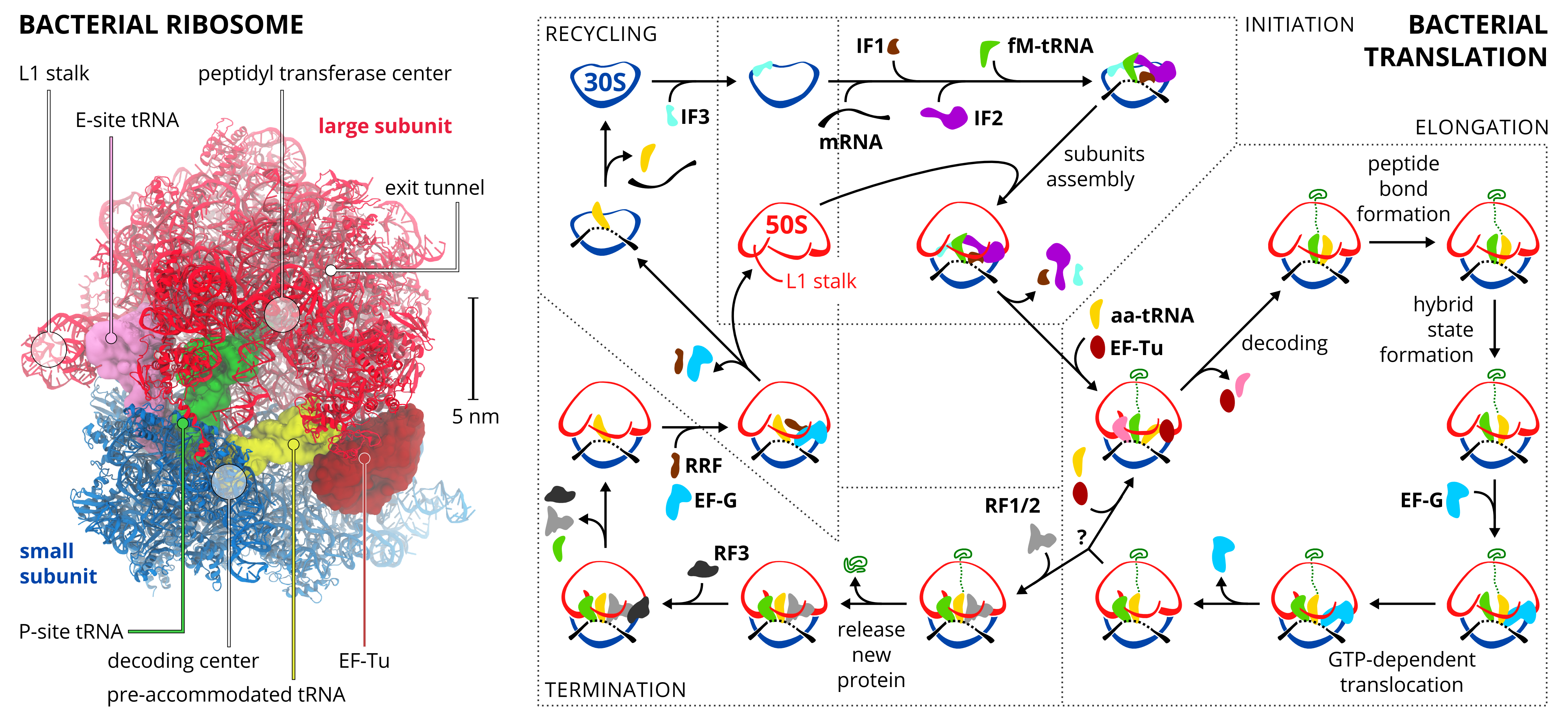}
\caption{ Structure of the bacterial ribosome in complex with EF-Tu (PDB 5AFI \cite{Fischer2015}). Scheme of the bacterial translation cycle as reviewed in Ref. \citenum{Schmeing2009a}. 30S: small subunit; 50S: large subunit; IF1, IF2, IF3: initiation factors; fM-tRNA: N-formylmethionine tRNA; aa-tRNA: aminoacyl tRNA; EF-Tu, EF-G: elongation factors; RF1, RF2, RF3: release factors; RRF: ribosome recycling factor; green trace: nascent protein. The question mark stands for a stop codon recognition.}
\label{fig:intro}
\end{center}
\end{figure*}

Ribosomes are large RNA-protein complexes which synthesize proteins in a process called translation \cite{Rodnina2007}. Translation proceeds in a multi-step cycle and involves a messenger RNA (mRNA), transfer RNAs (tRNAs) and a number of proteins (translation factors) (Figure \ref{fig:intro}). The investigation of ribosomes not only helps to understand protein synthesis and its regulation, but also offers a tremendous potential for medicinal applications. Indeed, the ribosome is one of the main antibacterial drug targets, and also the main actor in the problem of drug resistance \cite{Wilson2013}.

Advances in X-ray crystallography and cryo-EM provided a wealth of structural information about ribosomes at atomic detail \cite{Schmeing2009a,Frank2017}. 
However, the accessible information on ribosome dynamics, which is essential to understand ribosome function  (Figure 1), is limited due to preconditions of the experimental techniques. Specifically, high structural resolution in X-ray crystallography is obtained only for conformationally homogeneous crystals, and high-resolution cryo-EM requires many images of the complex in the same chemical and conformational state.

Molecular dynamics (MD) simulations have evolved into a powerful technique that complements the structural information. Based on first principles physics laws, it allows performing \textit{in silico} experiments, e.g., a removal of a chemical lock in the stalled ribosome \cite{Arenz2016}, that are inaccessible by other means. Obtaining free energies and transition rates from simulations allows a direct comparison to kinetic experiments, which is crucial for validation of the simulations and the proposed molecular mechanisms. Further, MD simulations biased by cryo-EM density maps have been successfully used as a tool for high-resolution structure determination (see Ref. \citenum{Kirmizialtin2015} and references therein). Finally, and possibly most importantly, MD simulations have enabled us to move from mere correlations to an understanding of causes and effects.

The MD field underwent a respectable progress since the first all-atom MD simulation of the entire ribosome \cite{Sanbonmatsu2005}. However, MD simulations of the ribosome still remain challenging for two main reasons: the large size of the ribosome and the wide range of time scales relevant to its function. To this aim, the whole arsenal of simulation  methods is used, ranging from coarse-grained MD simulations (cgMD) of the entire ribosome \cite{ Nilsson15, Deeng2016}, through structure-based MD (sbMD) \cite{Noel2016,Nguyen2016}, explicit-solvent all-atom MD simulations (aaMD) of entire ribosome \cite{Sanbonmatsu2005, Whitford2013a, Bock2013a, Ishida2014, Bock2015, Arenz2016, Cabrita16} or its reduced/cutout model \cite{Small2013, Lind2013, Wallin2013, Panecka2014, Satpati2014, Sothiselvam2014, Zeng2014, Makarov2015, Fischer2016, Lind2016, Huter2017}, to a quantum mechanical/molecular mechanical setup (QM/MM) explicitly treating the quantum character of electrons \cite{Xu2012, Aqvist2015, Swiderek2015}. The more detailed the description of the system, the higher is the computational effort, which leads to shorter accessible time scales. Due to the limited extent of this review, we recommend Refs. \citenum{Vangaveti2017, Sponer2017} for a more in depth discussion of these fundamental topics.

At least three reviews have been published \cite{Sanbonmatsu2012, Aqvist2012, Makarov2016} with similar scope to this Opinion. Here, we follow up previous reviews by Sanbonmatsu \cite{Sanbonmatsu2012} and \AA qvist et al.\ \cite{Aqvist2012} from 2012. Our review is organized according to the ribosomal translation cycle. We also cover two topics which are closely related to translation and where simulations have proven useful, namely the action of ribosome-binding antibiotics and cotranslational folding. Due to limited space, we can only discuss a fraction of the many articles published recently.

\section*{Initiation}

Translation is initialized on the AUG start codon on the mRNA. In eukaryotes, this codon is recognized by a pre-initiation complex of the small subunit, initiator tRNA and several initiation factors (IFs), which scans the mRNA \cite{Hinnebusch2014}. Not much simulation work has focused on initiation so far. One exception is a work of Lind and \AA qvist who investigated the role of IFs on codon recognition \cite{Lind2016}. Using aaMD they calculated relative binding free energies for single-point mutations of the start codon in the presence and absence of two IFs. The simulations suggest that the presence of the IFs on the pre-initiation complex increases the energetic penalty for binding non-cognate codons and, thereby, the fidelity of cognate start codon recognition is enhanced.

\section*{Decoding}
A large fraction of simulation work has focussed on the decoding step preceding peptide elongation, presumably because many important details of decoding do not involve large-scale conformational rearrangements. During elongation, aminoacyl-tRNAs are delivered to the ribosome in the form of a ternary complex: the tRNA, a translational GTPase (in bacteria: EF-Tu or SelB), and a GTP molecule. The tRNA decodes the information on the mRNA by forming hydrogen bonds (H-bonds) between codon and anticodon nucleobases. Remarkably, the free-energy difference between correct (cognate) and incorrect (near-cognate, non-cognate) base pairing alone does not explain the very high fidelity of decoding \cite{Rodnina2001}. Rather, high fidelity is achieved by a two-step decoding process: (i) initial selection leading to GTPase activation and (ii) proofreading. In addition to the free-energy difference, kinetic effects contribute to the discrimination. The GTP hydrolysis rate is increased and tRNA rejection rate is decreased by the recognition of the correct codon \cite{Rodnina2001}. 

Free-energy aaMD simulations of the decoding region were used to investigate the discrimination between near-cognate and cognate base pairs \cite{Satpati2014}. Small-subunit nucleotides A1492 and A1493 adopt a flipped-out conformation in the presence of a tRNA and, in this conformation, interact with the codon-anticodon mini-helix. In the simulations, flipped out nucleotides A1492, A1493 along with G530 were found to shield the codon-anticodon base pairs from solvent. This shielding prevents interactions of near-cognate base pairs with the solvent, thereby increasing the free-energy difference between near-cognate and cognate base pairs and thus the discrimination \cite{Satpati2014}. The shielding mechanism is consistent with X-ray structures of near-cognate complexes which imply that the decoding center enforces a base-pair geometry of mismatched base pairs close to that of a canonical base pair \cite{Demeshkina2012}. This leads to a reduction of possible H-bonds relative to a relaxed conformation.

Umbrella sampling aaMD of the in- and out-flipping of A1492/A1493 in a simulation system consisting of the decoding center suggested a more active role of these nucleotides \cite{Zeng2014}. In the simulations, a flipped out conformation was seen to be more favourable for cognate than for near-cognate base pairs, which would increase the discrimination by stabilizing the cognate codon-anticodon helix.

Aminoglycosides are a class of antibiotics that bind to the decoding center and lock nucleotides A1492/A1493 in the flipped-out conformation. In this way aminoglycosides promote the accommodation of near-cognate, thus wrong, tRNAs. Panecka et al.\ carried out  aaMD of the decoding center with bound aminoglycoside paromomycin and resistance-inducing mutations in protein uS12 \cite{Panecka2014}. An increased flipping rate relative to the wild type was observed, suggesting that the mutation restores the function of these nucleotides.

In a recent cryo-EM study, intermediate structures along the pathway of initial selection leading to the GTPase activation of SelB was resolved at high resolution \cite{Fischer2016}. Both tRNA and SelB undergo substantial conformational changes during the process which could present a kinetic barrier controlling GTPase activation. To address the question if these conformational changes by themselves are rate-limiting to the process, aaMD of the free ternary complex in solution was performed. The simulations were started from the ribosome-bound cryo-EM conformations. The tRNA and SelB rapidly interconverted between the different conformations, which allowed the construction of the conformational free-energy landscape. The landscape indicates that the intrinsic large-scale conformational changes of the tRNA and SelB during the delivery to the ribosome are not rate-limiting to the process.

In the GTPase-activated state, the GTP binding domain of EF-Tu or SelB docks onto the sarcin-ricin loop of the large subunit. Wallin et al.\ used aaMD of the GTP binding site of EF-Tu in the activated state on the ribosome to study the GTP hydrolysis \cite{Wallin2013}. Free-energy perturbation simulations suggested that His81 of EF-Tu cannot act as a general base in the reaction, as was previously proposed, but rather stabilizes a water molecule involved in the reaction. Further, the sarcin-ricin loop seems to promote the GTPase activated conformation of a conserved tripeptide motif PGH which contains His81. The conformation of the PGH and Mg$^\textrm{2+}$ positions predicted from these simulations were later confirmed by high-resolution X-ray structures (see references within \cite{Aqvist2015}). aaMD in combination with empirical valence bond (EVB) method suggested that His81 is doubly protonated which promotes a proton transfer from the stabilized water molecule to the GTP $\gamma$-phosphate, followed by a nucleophilic attack of the hydroxide ion \cite{Aqvist2015}. By calculating Arrhenius plots from MD simulations at multiple temperatures $\textrm{\AA}$qvist et al.\ identified a large entropic contribution to the hydrolysis reaction \cite{Aqvist2015a}.

After GTP hydrolysis, the GTPase dissociates from the tRNA which allows the tRNA to move into to the peptidyl transferase center (PTC) on the large subunit. A tRNA accommodation corridor was identified using targeted aaMD of the entire ribosome \cite{Sanbonmatsu2005}. aaMD and  sbMD of tRNA accommodation suggested that there is an additional intermediate state between the pre-accommodation (A/T) and the fully accommodated states \cite{Sanbonmatsu2005, Noel2016}. It is characterized by  the tRNA elbow in the accommodated state and the CCA-tail not yet in the PTC. Recent sbMD suggests that EF-Tu sterically reduces the range of accessible tRNA conformations, specifically in the A/T state \cite{Noel2016}. Therefore, the presence of EF-Tu is predicted to destabilize the A/T state and thereby enhance the rate of accommodation.

\section*{Peptide Bond Formation}

At the core of ribosomal translation is the catalysis of peptide bond formation \cite{Rodnina2007}. The current reaction models point to a substrate assisted mechanism. Early aaMD with EVB calculations suggested that the ribosome reduces the solvent reorganization energy by providing a stable H-bond network, which would enhance the peptide bond formation rates \cite{Trobro2005}. 

MD simulations identified positions of several water molecules and H-bonds critical for the reaction which were subsequently confirmed by X-ray structures \cite{Schmeing2005, Trobro2006}. QM/MM simulations as well as high-level quantum chemical calculations indicated that the transition state forms an eight membered ring which includes a water molecule and that the C--O bond cleavage takes place after C--N bond formation \cite{Wallin2010, Xu2012}. These studies reproduced the experimentally observed catalytic effect. A recent QM/MM study additionally proposed the presence of a Mg$^\textrm{2+}$ ion in the surrounding of the PTC and has shown that including the ion improved agreement of the calculated with the measured catalytic effect, underscoring the importance of ions in computational studies of the ribosome \cite{Swiderek2015}.

\begin{figure*}
\begin{center}
\includegraphics[width=0.8\textwidth]{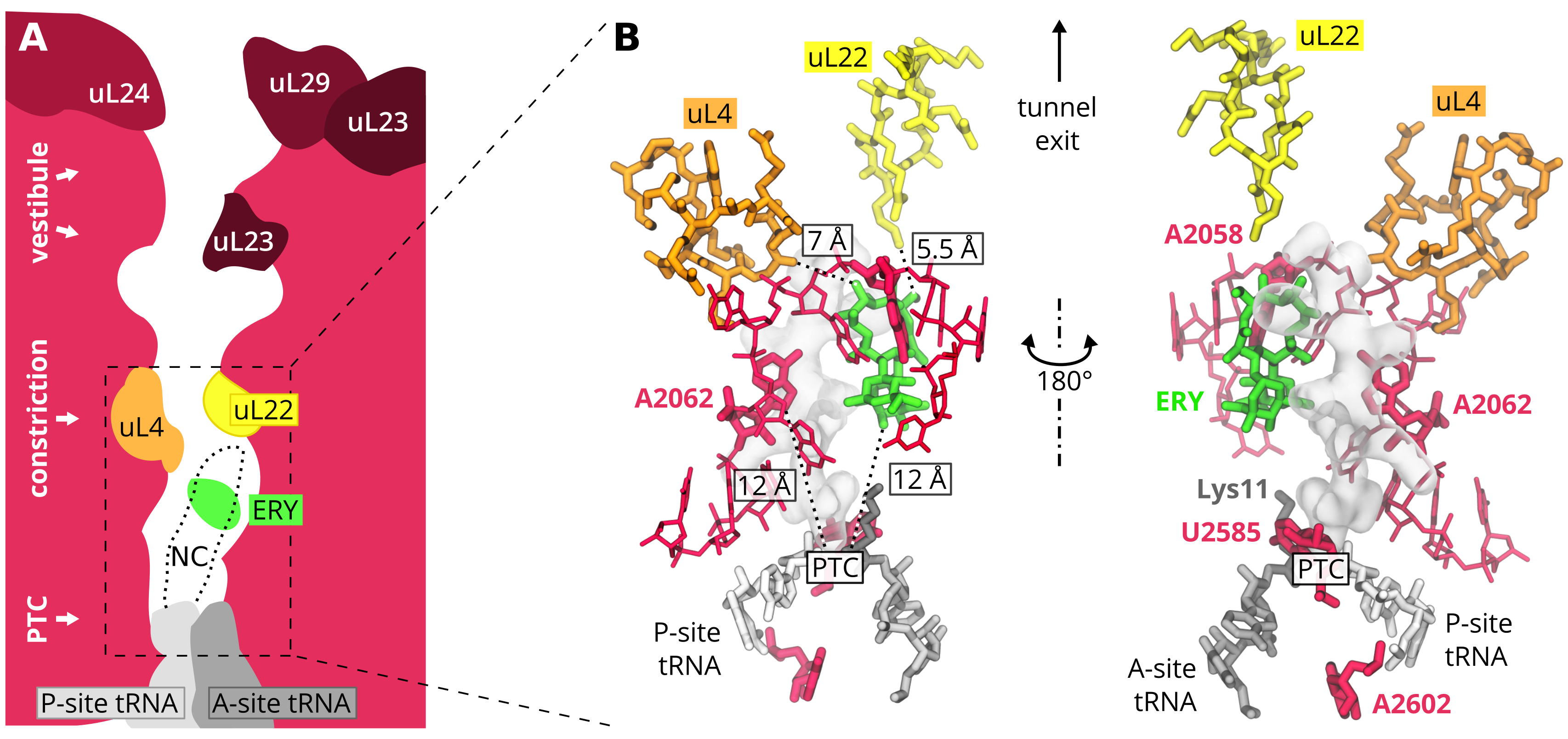}
\caption{A: Scheme of the ribosome exit tunnel with several proteins highlighted. NC, nascent chain; ERY, erythromycin; PTC, peptidyl transferase center. B: Context of the erythromycin (ERY, in green) binding; figure based on PDB: 5JTE \cite{Arenz2016}. Several large subunit nucleotides are highlighted in bold red. Two proteins uL4 and uL22 form a constriction site. The nascent peptide is shown as transparent surface.}
\label{fig:tunnel}
\end{center}
\end{figure*}

Since the PTC is buried within the large subunit, during translation the nascent chain (NC) exits through a 100-\AA ~long tunnel (Figure \ref{fig:tunnel}A). The exit tunnel plays an active role in protein synthesis. Certain peptide sequences specifically interact with tunnel walls and induce ribosome stalling \cite{Wilson2011}. Further, the exit tunnel is a binding site for a clinically important class of antibiotics \cite{Wilson2013}. Several MD studies have addressed these issues.

When synthesizing proteins containing proline stretches (i.e. several prolines in a row), ribosomes  become stalled. Stalling is alleviated by a specialized elongation factor, EF-P in bacteria \cite{Doerfel2013,Ude2013}.
Recently, cryo-EM structures of a ribosome stalled by a proline stretch with and without EF-P were resolved \cite{Huter2017}. In aaMD simulations of the PTC region, EF-P was observed to stabilize the P-site tRNA in a conformation compatible with peptide bond formation, while in absence of EF-P, the tRNA moved away from the A-site tRNA \cite{Huter2017}. 


A communication pathway between the tunnel walls and the PTC was identified in antibiotic-dependent ribosome stalling. Biochemical experiments complemented by aaMD of the entire \textit{E.\ coli} ribosome suggested \cite{Sothiselvam2014} that a macrolide antibiotic erythromycin (ERY) allosterically alters the properties of the PTC without any direct contact with the NC (Figure 2). The simulations  captured a dramatic reorientation of U2585 and A2602, both more than 8 $\textrm{\AA}$ away from the nearest ERY atom. Beside this, ERY binding induced a conformational change of its neighbor A2062, a nucleotide claimed to serve as a nascent-chain sensor \cite{Vazquez-Laslop2010}.

The role of A2062 was further highlighted by aaMD of a cubic reduced model of the PTC and its surroundings in the presence/absence of ERY \cite{Makarov2015}. Over twenty unbiased trajectories, each 200-360 ns, suggested that allosteric signal transmission occurs via formation of a stem of stacked large-subunit ribosomal RNA nucleobases and that, further, the stem formation is initiated by the A2062 conformational change.

ERY-induced stalling of an ErmB leading peptide (ErmBL) \cite{Arenz2016} was studied using  aaMD of the entire \textit{E.\ coli} ribosome with ErmBL peptide in the presence and absence of ERY, complementing cryo-EM experiments which can not resolve unstalled ribosome-ErmBL complexes (without ERY). The simulations suggested that the ERY induced conformational changes in the PTC result in a shift of a Lys11 on the A-site tRNA, thereby preventing peptide bond formation. The simulations predicted a mutation of the Lys11 to enhance the stalling ability of ErmBL, which was subsequently confirmed by biochemical experiments \cite{Arenz2016}.

The resistance to a ketolide antibiotic telithromycin (TEL) was addressed by MacKerell's group \cite{Small2013}. aaMD of a spherical reduced model were combined with grand canonical Monte Carlo to enhance sampling of water positions in the exit tunnel. The study compared the wild-type complex to three drug-resistant variants: A2058G mutant as well as mono- and dimethylated A2058. Based on geometric analyses of H-bonds and other contacts it was rationalized why TEL mitigates the drug resistance due to A2058G mutation but remains susceptible to methyl-mediated resistance. The key player seems to be an H-bond between TEL:2'-OH and A2058 that is maintained after mutation to G2058, but disrupted by methylations.

\section*{Cotranslational Protein Folding}

The exit tunnel can accommodate 30--60 AAs, depending on the level of NC compaction. The rate of translation of about 4--22 AA per second in bacteria \cite{Wohlgemuth2011} provides the NC with sufficient time to explore its conformational space and to start folding when still bound to the ribosome-tRNA complex. A number of simulation studies has tackled cotranslational folding (See Ref. \citenum{Trovato16} for a dedicated review). Particular interest lies in the trigger factor (TF), the first chaperone encountered by the NC.

Free TF in solution exhibits a monomer-dimer equilibrium. It was studied by several groups using aaMD \cite{Thomas2013, Singhal2013, Can2017}. The TF's N-terminal and head domains can associate making a compact tertiary structure. The conclusion that TF is very flexible is supported across all the studies, however the stability of the compact structure is force-field dependent and remains elusive. It was pointed out though that the compact structure prevents TF from binding to the ribosome, as supported by an elastic network model of the large subunit and TF complex \cite{Can2017}.

The association of TF and \textit{E.\ coli} ribosome was studied by cryo-EM and cgMD of a reduced ribosome model \cite{Deeng2016}. Several 1.2-${\mu}$s long trajectories showed interdomain motions: while the N-terminal domain remained bound to the ribosome, the head domain fluctuated between bound and unbound states. The motions were similar to those identified in solution simulations \cite{Thomas2013, Singhal2013}. In the ribosome-bound simulations, longer NC made the TF more rigid as compared to the shorter NC construct. The authors speculated that the loss of TF flexibility may facilitate TF unbinding from the ribosome, while keeping TF and NC still together.

Using NMR and cgMD, Deckert et al.\ studied the role of TF in synthesis of the disordered peptide $\alpha$-synuclein ($\alpha$Syn) \cite{Deckert2016}. Decent agreement between NMR and cgMD was observed. The authors suggested weak association of $\alpha$Syn and ribosome surface and concluded that there might be a specific affinity on the surface for aromatic residues. About 50 AAs are needed from the PTC to initiate interactions with TF.

NMR was used to study cotranslational folding of a pair of immunoglobulin-like domains FNL5 and FLN6 \cite{Cabrita16}. aaMD restrained by chemical shifts generated an ensemble of protein conformations on the ribosome. The N-terminal FNL5 domain was shown to adopt native-like fold only after emerged well beyond the tunnel. The FNL6, which was located closer to the C-terminus than the FNL5, was disordered yet compact and transiently interacted with the ribosomal surface.

A combined simulation-experimental study showed that a protein can fold already inside the tunnel vestibule (Figure \ref{fig:tunnel}A) \cite{Nilsson15}. A zinc-finger domain, probed by cgMD, cryo-EM, and biochemical experiments on stalled ribosomes, folds in the tunnel between uL22 and uL23 proteins. The folding was observed in cgMD at physiological (310 K) as well as at cryo-EM (140 K) temperatures, although the structural agreement with the cryo-EM model was rather poor.

\section*{tRNA Translocation}
After peptide bond formation the NC is attached to the tRNA residing in the A site, the tRNA in the P site is left deacylated and the E site is unoccupied (Figure 3a). The two tRNAs then translocate to the P and E sites, either spontaneously on slow time scales, or very rapidly in the presence of the translational GTPase EF-G. Translocation of tRNAs is accompanied by large-scale collective motions of the ribosome: relative rotation of ribosomal subunits and L1-stalk motion. The L1 stalk, which is a flexible part of the large subunit, is in contact and moves along with the tRNA from the P to the E site (Figure 3b).

Whitford et al.\ extracted effective diffusion coefficients for small subunit head and body rotations and tRNA displacement from an aaMD of the classical pre-translocation state \cite{Whitford2013a}. The diffusion coefficients together with experimental rates of these motions provide upper-bound estimates of free-energy barriers.

To obtain dynamics and energetics of intermediate states of spontaneous translocation, X-ray structures were refined against cryo-EM reconstructions, thereby obtaining 13 near-atomic resolution structures \cite{Bock2013a}. From aaMDs of these intermediate states, order-of-magnitude transition rates between states were estimated for motions of the L1 stalk and the tRNAs as well as for intersubunit rotations (Figure 3c). These rates revealed rapid dynamics of the L1 stalk and intersubunit rotations on sub-microsecond timescales, whereas the tRNA motions were seen to be rate-limiting for most transitions. Further, calculated molecular forces revealed that the L1 stalk is actively pulling the tRNA from P to E site as one of the main mechanisms accelerating barrier crossing.

\begin{figure*}
\begin{center}
\includegraphics[width=\textwidth]{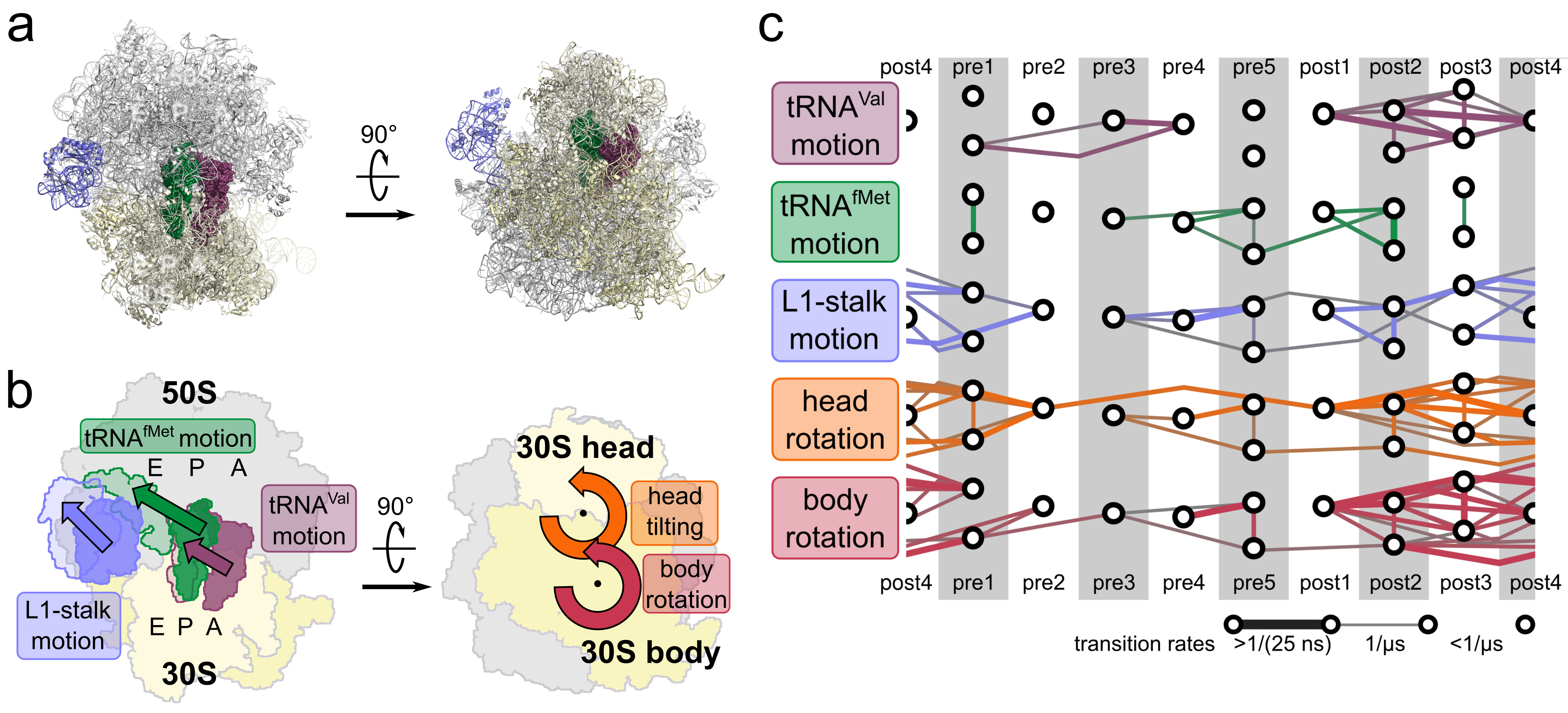}
\caption{ (a) Pre-translocation structure of the ribosome with tRNAs in A and P sites (purple, green). (b) Motions accompanying tRNA translocation. (c) Transition rates for different motions between different states estimated from the simulations. Figure adapted from Ref. \citenum{Bock2013a}.}
\label{fig:transl}
\end{center}
\end{figure*}

Using aaMD, it was shown that the intersubunit contact network adapts to different intersubunit rotation angles resulting in a relatively constant intersubunit binding enthalpy \cite{Bock2015}. A rotation-independent affinity between the subunit is a prerequisite for rapid rotation since different affinities would lead to barriers hindering rotation. In addition to the central contacts close to the axis of rotation, peripheral contacts were found to be strong, despite their large relative shift. This steady contribution is achieved by exchanging contact partners in the course of rotation.

Recently, sbMD simulations were applied to study the relation between small subunit head rotation and the mRNA-tRNA translocation \cite{Nguyen2016}. The authors chose the non-rotated state of the ribosome as an energy minimum as well as two adjacent binding sites for each of the two tRNAs. The model was able to reproduce a known intermediate tRNA binding conformations (ap/P--pe/E) and predicted an intermediate state with a tilted head, which has not been observed in experiments yet. In this proposed intermediate, the PE loop, which sterically separates the P and E binding sites on the small subunit, is displaced, possibly allowing the anticodon stem loop of the tRNA to move from P to E site. After removal of steric interactions of tRNA with small subunit protein uS13 and the PE loop, the tilted head state was not populated indicating that tilting results from these  interactions.

In an earlier study, Ishida et al.\ applied a cryo-EM fitting MD approach to study reverse translocation in the presence of EF-G \cite{Ishida2014}. The simulations were started from a post-translocation structure and then driven towards a pre-translocation state by maximising the correlation to a pre-translocation cryo-EM map. A free-energy landscape was estimated using umbrella simulations. Unexpectedly, they observed a clockwise rotation in the simulations, when going from the post state to the intermediate state, while from cryo-EM structures of intermediate states a counter-clockwise rotation was expected. The authors speculated that the movement of the head during reverse translocation might be different from that in what they call ordinary translocation. However, since the simulations describe an equilibrium process, there can be no difference due to directionality. The tilted head state described by Nguyen et al.\ \cite{Nguyen2016} was not observed in the simulations by Ishida et al. The differences between the observed free-energy landscapes underscore the severe sampling problem which is a major challenge for all simulations of such large molecular complexes.

\section*{Termination}
The translation cycle terminates in a series of steps after an mRNA stop codon is presented in the A site (Figure 1). First, a release factor (RF) recognizes a stop codon and hydrolyses the peptidyl-tRNA bond in the PTC. In the last step of termination, the large and small subunits of the ribosome separate. The subunits are recycled in the next translation round.

In mitochondria, non-standard stop codons have evolved \cite{Garcia-Guerrero2016}. In vertebrates, it is still unclear what factors recognize stop codons and how many stop codons actually exist. Lind et al.\ carried out aaMD of a reduced spherical model and calculated relative binding free energies of several codons when bound to a variety of RFs 
\cite{Lind2013}. Their results suggest that neither of the two mitochondrial RFs mtRF1 and mtRF1a, homologs to the bacterial RF1, is able to read non-standard AGA nor AGG stop codons. The authors advocated for another factor - ICT1 - that could be responsible for the peptide release in mitochondria as supported by later biochemical experiments \cite{Akabane2014}.

The hydrolysis of peptidyl-tRNA bond allows releasing the newly synthesized protein from the ribosome. The atomistic details of the hydrolysis mechanism are not yet fully understood. Kazemi et al.\ tested several mechanisms by means of density functional theory on a reduced model of the bacterial PTC (224 atoms) \cite{Kazemi2016}. They concluded that a base-catalyzed mechanism, which involves a deprotonation of the P-site tRNA A76 2'-OH group, is the only one consistent with the experimental activation energies, kinetic solvent isotope effect values, and pH dependence.

\section*{Summary and Outlook}
An increasing number of theory and computation research groups have accepted the challenges posed by a macromolecular system of the huge size of a ribosome, yielding a wealth of mechanistic insights into this amazing molecular machine. This advance has been fueled by the 'resolution revolution' of the cryo-EM field, and the obtained high-resolution structures of many conformational states. Most of the striking results in this field have not been obtained by isolated simulation work, but in close collaborations with other fields, most notably cryo-EM, spectroscopy, and biochemistry.

In this context, we see a few avenues which have, in our view, not yet been exploited to their full potential by simulations. First, the new high-resolution cryo-EM maps obtained at high pace demand improved and automated refinement protocols, where MD may play a vital role. Second, realistic simulations of single-molecule experiments will enhance the structural and dynamic interpretation. And, finally, more reliable protocols and methods for accurate free-energy calculations are key to a quantitative and causal understanding of how ribosomes work. We think such an understanding will have to be in terms of molecular driving forces, Markov processes, and ultimately the full dynamics within a multidimensional free-energy landscape. Understanding the ribosome remains, therefore, not only a computational and methodological, but also a conceptual challenge.

\section*{Acknowledgement}
The work was supported by the Max Planck Society and by the Deutsche Forschungsgemeinschaft (FOR 1805).


\footnotesize

\begin{thebibliography}{10}

\bibitem{Rodnina2007}
M.~V. Rodnina, M. Beringer, and W. Wintermeyer.
\newblock {How ribosomes make peptide bonds}.
\newblock {\em Trends Biochem. Sci.}, 32(1):20--6, 2007.

\bibitem{Wilson2013}
D.~N. Wilson.
\newblock {Ribosome-targeting antibiotics and mechanisms of bacterial
  resistance}.
\newblock {\em Nat. Rev. Microbiol.}, 12(1):35--48, 2014.

\bibitem{Fischer2015}
N. Fischer, P. Neumann, A.~L. Konevega, L.~V. Bock, R. Ficner, M.~V. Rodnina,
  and H. Stark.
\newblock {Structure of the E. coli ribosome--EF-Tu complex at {\textless}3
  {\AA} resolution by Cs-corrected cryo-EM}.
\newblock {\em Nature}, 520:567--570, 2015.

\bibitem{Schmeing2009a}
T.~M. Schmeing and V. Ramakrishnan.
\newblock {What recent ribosome structures have revealed about the mechanism of
  translation}.
\newblock {\em Nature}, 461(7268):1234--1242, 2009.

\bibitem{Frank2017}
J. Frank.
\newblock {The translation elongation cycle---capturing multiple states by
  cryo-electron microscopy}.
\newblock {\em Philos. Trans. R. Soc. Lon. B., Biol. Sci.}, 372(1716), 2017.

\bibitem{Arenz2016}
S. Arenz, L.~V. Bock, M. Graf, C.~A. Innis, R. Beckmann, H. Grubm{\"{u}}ller,
  A.~C. Vaiana, and D.~N. Wilson.
\newblock {A combined cryo-EM and molecular dynamics approach reveals the
  mechanism of ErmBL-mediated translation arrest}.
\newblock {\em Nat. Commun.}, 7:12026, 2016.\\
$\bullet\bullet$ All-atom MD simulations of the entire ribosome together with cryo-EM reveal molecular details of an antibiotic-dependent ribosomal stalling mechanism which are confirmed by toeprinting experiments. This study emphasizes the benefits of an interdisciplinary approach.

\bibitem{Kirmizialtin2015}
S. Kirmizialtin, J. Loerke, E. Behrmann, C.~M.~T. Spahn, and K.~Y. Sanbonmatsu.
\newblock {Using molecular simulation to model high-resolution cryo-EM
  reconstructions}.
\newblock {\em Methods Enzymol.}, 558(1):497--514, 2015.

\bibitem{Sanbonmatsu2005}
K.~Y. Sanbonmatsu, S. Joseph, and C.-S. Tung.
\newblock {Simulating movement of tRNA into the ribosome during decoding}.
\newblock {\em Proc. Natl. Acad. Sci. U. S. A.}, 102(44):15854--15859, 2005.

\bibitem{Nilsson15}
O.~B. Nilsson, R. Hedman, J. Marino, S. Wickles, L. Bischoff, M. Johansson, A.
  M{\"u}ller-Lucks, F. Trovato, J.~D. Puglisi, E.~P. O’Brien, R. Beckmann,
  and G. von Heijne.
\newblock Cotranslational protein folding inside the ribosome exit tunnel.
\newblock {\em Cell Rep.}, 12(10):1533--1540, 2015.

\bibitem{Deeng2016}
J. Deeng, K.~Y. Chan, E.~O. van~der Sluis, O. Berninghausen, W. Han, J.
  Gumbart, K. Schulten, B. Beatrix, and R. Beckmann.
\newblock {Dynamic Behavior of Trigger Factor on the Ribosome}.
\newblock {\em J. Mol. Biol.}, 428(18):3588--3602, 2016.\\
$\bullet$ Cryo-EM study is complemented by coarse-grained MD simulations of the ribosome with trigger factor. They authors propose a way, how the trigger factor interacts with the emerging nascent chain.

\bibitem{Noel2016}
J.~K. Noel and P.~C. Whitford.
\newblock {How EF-Tu can contribute to efficient proofreading of aa-tRNA by the
  ribosome}.
\newblock {\em Nat. Commun.}, 7:13314, 2016.\\
$\bullet$  A unique combination of structure-based MD of a ribosome EF-Tu complex and kinetic modelling suggested that the presence of EF-Tu enhances the rate of tRNA accommodation.

\bibitem{Nguyen2016}
K. Nguyen and P.~C. Whitford.
\newblock {Steric interactions lead to collective tilting motion in the
  ribosome during mRNA-tRNA translocation}.
\newblock {\em Nat. Commun.}, 7:10586, 2016.\\
$\bullet$ Structure-based MD simulations of spontaneous tRNA translocation, including and excluding certain interactions, indicate that the small subunit head tilting is a consequence of tRNA-mRNA interactions. The possibility to switch off interactions computationally helps revealing causality.

\bibitem{Whitford2013a}
P.~C. Whitford, S.~C. Blanchard, J.~H.~D. Cate, K.~Y. Sanbonmatsu, and H. Lee.
\newblock {Connecting the Kinetics and Energy Landscape of tRNA Translocation
  on the Ribosome}.
\newblock {\em {PLoS} Comp. Biol.}, 9(3):e1003003, 2013.

\bibitem{Bock2013a}
L.~V. Bock, C. Blau, G.~F. Schr{\"{o}}der, I.~I. Davydov, N. Fischer, H. Stark,
  M.~V. Rodnina, A.~C. Vaiana, and H. Grubm{\"{u}}ller.
\newblock {Energy barriers and driving forces in tRNA translocation through the
  ribosome}.
\newblock {\em Nat. Struct. Mol. Biol.}, 20(12):1390--6, 2013.

\bibitem{Ishida2014}
H. Ishida and A. Matsumoto.
\newblock {Free-energy landscape of reverse tRNA translocation through the
  ribosome analyzed by electron microscopy density maps and molecular dynamics
  simulations}.
\newblock {\em Plo{S} {ONE}}, 9(7):e101951, 2014.

\bibitem{Bock2015}
L.~V. Bock, C. Blau, A.~C. Vaiana, and H. Grubm{\"{u}}ller.
\newblock {Dynamic contact network between ribosomal subunits enables rapid
  large-scale rotation during spontaneous translocation}.
\newblock {\em Nucleic Acids Res.}, 43(14):6747--60, 2015.\\
$\bullet$ All-atom MD of entire ribosome is used to simulate 13 conformational states. Combining the dynamics obtained for the different states, the study suggests intersubunit interactions have evolved to enable rapid large-scale rotation.

\bibitem{Cabrita16}
L.~D. Cabrita, A.~M. Cassaignau, H.~M. Launay, C.~A. Waudby, T. Wlodarski, C.
  Camilloni, M.-E. Karyadi, A.~L. Robertson, X. Wang, A.~S. Wentink, L.~S.
  Goodsell, C.~A. Woodhead, M. Vendruscolo, C.~M. Dobson, and J. Christodoulou.
\newblock A structural ensemble of a ribosome-nascent chain complex during
  cotranslational protein folding.
\newblock {\em Nat. Struct. Mol. Biol.}, 23(4):278--285, 2016.

\bibitem{Small2013}
M.~C. Small, P. Lopes, R.~B. Andrade, and A.~D. MacKerell.
\newblock {Impact of Ribosomal Modification on the Binding of the Antibiotic
  Telithromycin Using a Combined Grand Canonical Monte Carlo/Molecular Dynamics
  Simulation Approach}.
\newblock {\em {PLoS} Comp. Biol.}, 9(6):e1003113, 2013.

\bibitem{Lind2013}
C. Lind, J. Sund, and J. {\AA}qvist.
\newblock {Codon-reading specificities of mitochondrial release factors and
  translation termination at non-standard stop codons}.
\newblock {\em Nat. Commun.}, 4:2940, 2013.

\bibitem{Wallin2013}
G. Wallin, S.~C.~L. Kamerlin, and J. {\AA}qvist.
\newblock {Energetics of activation of GTP hydrolysis on the ribosome}.
\newblock {\em Nat. Commun.}, 4:1733, 2013.

\bibitem{Panecka2014}
J. Panecka, C. Mura, and J. Trylska.
\newblock {Interplay of the Bacterial Ribosomal A-Site, S12 Protein Mutations
  and Paromomycin Binding: A Molecular Dynamics Study}.
\newblock {\em PloS one}, 9(11):e111811, 2014.

\bibitem{Satpati2014}
P. Satpati, J. Sund, and J. {\AA}qvist.
\newblock {Structure-based energetics of mRNA decoding on the ribosome}.
\newblock {\em Biochemistry}, 53(10):1714--1722, 2014.

\bibitem{Sothiselvam2014}
S. Sothiselvam, B. Liu, W. Han, H. Ramu, D. Klepacki, G.~C. Atkinson, A.
  Brauer, M. Remm, T. Tenson, K. Schulten, N. V{\'{a}}zquez-Laslop, and A.~S.
  Mankin.
\newblock {Macrolide antibiotics allosterically predispose the ribosome for
  translation arrest}.
\newblock {\em Proc. Natl. Acad. Sci. U. S. A.}, 111(27):9804--9, 2014.

\bibitem{Zeng2014}
X. Zeng, J. Chugh, A. Casiano-Negroni, H.~M. Al-Hashimi, and C.~L. Brooks.
\newblock {Flipping of the Ribosomal A-Site Adenines Provides a Basis for tRNA
  Selection}.
\newblock {\em J. Mol. Biol.}, 426(19):3201--3213, 2014.

\bibitem{Makarov2015}
G.~I. Makarov, A.~V. Golovin, N.~V. Sumbatyan, and A.~A. Bogdanov.
\newblock {Molecular dynamics investigation of a mechanism of allosteric signal
  transmission in ribosomes}.
\newblock {\em Biochem. Mosc.}, 80(8):1047--1056, 2015.

\bibitem{Fischer2016}
N. Fischer, P. Neumann, L.~V. Bock, C. Maracci, Z. Wang, A. Paleskava, A.~L.
  Konevega, G.~F. Schr{\"{o}}der, H. Grubm{\"{u}}ller, R. Ficner, M.~V.
  Rodnina, and H. Stark.
\newblock {The pathway to GTPase activation of elongation factor SelB on the
  ribosome}.
\newblock {\em Nature}, 540:80--85, 2016.

\bibitem{Lind2016}
C. Lind and J. {\AA}qvist.
\newblock {Principles of start codon recognition in eukaryotic translation
  initiation}.
\newblock {\em Nucleic Acids Res.}, 44(17):8425--8432, 2016.

\bibitem{Huter2017}
P. Huter, S. Arenz, L.~V. Bock, M. Graf, J.~O. Frister, A. Heuer, L. Peil,
  A.~L. Starosta, I. Wohlgemuth, F. Peske, J. Nov{\'{a}}{\v{c}}ek, O.
  Berninghausen, H. Grubm{\"{u}}ller, T. Tenson, R. Beckmann, M.~V. Rodnina,
  A.~C. Vaiana, and D.~N. Wilson.
\newblock {Structural Basis for Polyproline-Mediated Ribosome Stalling and
  Rescue by the Translation Elongation Factor EF-P}.
\newblock {\em Mol. Cell}, 68(3):515--527.e6, 2017.

\bibitem{Xu2012}
J. Xu, J.~Z.~H. Zhang, and Y. Xiang.
\newblock {Ab Initio QM/MM Free Energy Simulations of Peptide Bond Formation in
  the Ribosome Support an Eight-Membered Ring Reaction Mechanism}.
\newblock {\em J. Am. Chem. Soc.}, 134(39):16424--16429, 2012.

\bibitem{Aqvist2015}
J. {\AA}qvist and S.~C.~L. Kamerlin.
\newblock {The Conformation of a Catalytic Loop Is Central to GTPase Activity
  on the Ribosome}.
\newblock {\em Biochemistry}, 54(2):546--556, 2015.\\
$\bullet\bullet$ Extensive QM/MM simulations of GTP hydrolysis on EF-Tu give important insights into the reaction mechanism in line with kinetic experiments and, moreover, are able to rationalize the effects of mutations.

\bibitem{Swiderek2015}
K. {\'{S}}widerek, S. Marti, I. Tu{\~{n}}{\'{o}}n, V. Moliner, and J. Bertran.
\newblock {Peptide Bond Formation Mechanism Catalyzed by Ribosome}.
\newblock {\em J. Am. Chem. Soc.}, 137(37):12024--12034, 2015.

\bibitem{Vangaveti2017}
S. Vangaveti, S.~V. Ranganathan, and A.~A. Chen.
\newblock {Advances in RNA molecular dynamics: a simulator's guide to RNA force
  fields}, 2017.

\bibitem{Sponer2017}
J. {\v{S}}poner, M. Krepl, P. Ban{\'{a}}{\v{s}}, P. K{\"{u}}hrov{\'{a}}, M.
  Zgarbov{\'{a}}, P. Jure{\v{c}}ka, M. Havrila, and M. Otyepka.
\newblock {How to understand atomistic molecular dynamics simulations of RNA
  and protein--RNA complexes?}
\newblock {\em WIRES RNA}, 8(3):e1405, 2017.

\bibitem{Sanbonmatsu2012}
K.~Y. Sanbonmatsu.
\newblock {Computational studies of molecular machines: The ribosome}.
\newblock {\em Current Opinion in Structural Biology}, 22(2):168--174, 2012.

\bibitem{Aqvist2012}
J. {\AA}qvist, C. Lind, J. Sund, and G. Wallin.
\newblock {Bridging the gap between ribosome structure and biochemistry by
  mechanistic computations}.
\newblock {\em Curr. Opin. Struc. Biol.}, 22(6):815--23, 2012.

\bibitem{Makarov2016}
G.~I. Makarov, T.~M. Makarova, N.~V. Sumbatyan, and A.~A. Bogdanov.
\newblock {Investigation of ribosomes using molecular dynamics simulation
  methods}.
\newblock {\em Biochem. Mosc.}, 81(13):1579--1588, 2016.

\bibitem{Hinnebusch2014}
A.~G. Hinnebusch.
\newblock {The Scanning Mechanism of Eukaryotic Translation Initiation}.
\newblock {\em Ann. Rev. Biochem.}, 83(1):779--812, 2014.

\bibitem{Rodnina2001}
M.~V. Rodnina and W. Wintermeyer.
\newblock {Fidelity of Aminoacyl-tRNA Selection on the Ribosome: Kinetic and
  Structural Mechanisms}.
\newblock {\em Ann. Rev. Biochem.}, 70(1):415--435, 2001.

\bibitem{Demeshkina2012}
N. Demeshkina, L. Jenner, E. Westhof, M. Yusupov, and G. Yusupova.
\newblock {A new understanding of the decoding principle on the ribosome}.
\newblock {\em Nature}, 484(7393):256--259, 2012.

\bibitem{Aqvist2015a}
J. {\AA}qvist and S.~C.~L. Kamerlin.
\newblock {Exceptionally large entropy contributions enable the high rates of
  GTP hydrolysis on the ribosome}.
\newblock {\em Sci. Rep.}, 5:15817, 2015.

\bibitem{Trobro2005}
S. Trobro and J. {\AA}qvist.
\newblock {Mechanism of peptide bond synthesis on the ribosome}.
\newblock {\em Proc. Natl. Acad. Sci. U. S. A.}, 102(35):12395--400, 2005.

\bibitem{Schmeing2005}
T.~M. Schmeing, K.~S. Huang, D.~E. Kitchen, S.~A. Strobel, and T.~A. Steitz.
\newblock {Structural insights into the roles of water and the 2′ hydroxyl of
  the P site tRNA in the peptidyl transferase reaction}.
\newblock {\em Mol. Cell}, 20(3):437--448, 2005.

\bibitem{Trobro2006}
S. Trobro and J. {\AA}qvist.
\newblock {Analysis of predictions for the catalytic mechanism of ribosomal
  peptidyl transfer}.
\newblock {\em Biochemistry}, 45(23):7049--56, 2006.

\bibitem{Wallin2010}
G. Wallin and J. {\AA}qvist.
\newblock {The transition state for peptide bond formation reveals the ribosome
  as a water trap}.
\newblock {\em Proc. Natl. Acad. Sci. U. S. A.}, 107(5):1888--93, 2010.

\bibitem{Wilson2011}
D.~N. Wilson and R. Beckmann.
\newblock {The ribosomal tunnel as a functional environment for nascent
  polypeptide folding and translational stalling}.
\newblock {\em Current Opinion in Structural Biology}, 21(2):274--282, 2011.

\bibitem{Doerfel2013}
L.~K. Doerfel, I. Wohlgemuth, C. Kothe, F. Peske, H. Urlaub, and M.~V. Rodnina.
\newblock {EF-P is essential for rapid synthesis of proteins containing
  consecutive proline residues.}
\newblock {\em Science (New York, N.Y.)}, 339(6115):85--8, 2013.

\bibitem{Ude2013}
S. Ude, J. Lassak, A.~L. Starosta, T. Kraxenberger, D.~N. Wilson, and K. Jung.
\newblock {Translation elongation factor EF-P alleviates ribosome stalling at
  polyproline stretches.}
\newblock {\em Science (New York, N.Y.)}, 339(6115):82--5, 2013.

\bibitem{Vazquez-Laslop2010}
N. V{\'{a}}zquez-Laslop, H. Ramu, D. Klepacki, K. Kannan, and A.~S. Mankin.
\newblock {The key function of a conserved and modified rRNA residue in the
  ribosomal response to the nascent peptide}.
\newblock {\em EMBO J.}, 29(18):3108--3117, 2010.

\bibitem{Wohlgemuth2011}
I. Wohlgemuth, C. Pohl, J. Mittelstaet, A.~L. Konevega, and M.~V. Rodnina.
\newblock {Evolutionary optimization of speed and accuracy of decoding on the
  ribosome}.
\newblock {\em Philos. Trans. R. Soc. Lon. B., Biol. Sci.},
  366(1580):2979--2986, 2011.

\bibitem{Trovato16}
F. Trovato and E.~P. O'Brien.
\newblock Insights into cotranslational nascent protein behavior from computer
  simulations.
\newblock {\em Ann. Rev. Biophys.}, 45:345--369, 2016.

\bibitem{Thomas2013}
A.~S. Thomas, S. Mao, and A.~H. Elcock.
\newblock {Flexibility of the Bacterial Chaperone Trigger Factor in
  Microsecond-Timescale Molecular Dynamics Simulations}.
\newblock {\em Biophys. J.}, 105(3):732--744, 2013.

\bibitem{Singhal2013}
K. Singhal, J. Vreede, A. Mashaghi, S.~J. Tans, and P.~G. Bolhuis.
\newblock {Hydrophobic Collapse of Trigger Factor Monomer in Solution}.
\newblock {\em PLoS ONE}, 8(4):e59683, 2013.

\bibitem{Can2017}
M.~T. Can, Z. Kurkcuoglu, G. Ezeroglu, A. Uyar, O. Kurkcuoglu, and P. Doruker.
\newblock {Conformational dynamics of bacterial trigger factor in apo and
  ribosome-bound states}.
\newblock {\em PLoS ONE}, 12(4):e0176262, 2017.

\bibitem{Deckert2016}
A. Deckert, C.~A. Waudby, T. Wlodarski, A.~S. Wentink, X. Wang, J.~P.
  Kirkpatrick, J.~F.~S. Paton, C. Camilloni, P. Kukic, C.~M. Dobson, M.
  Vendruscolo, L.~D. Cabrita, and J. Christodoulou.
\newblock {Structural characterization of the interaction of $\alpha$-synuclein
  nascent chains with the ribosomal surface and trigger factor}.
\newblock {\em Proc. Natl. Acad. Sci. U. S. A.}, 113(18):5012--5017, 2016.\\
$\bullet\bullet$ A practical combination of NMR and MD simulations describes the cotranslational behavior of an intrinsically disordered peptide $\alpha$-synuclein. The simulations mimic a translating ribosome in which the nascent chain length gradually increases.	

\bibitem{Garcia-Guerrero2016}
A.~E. Garc{\'{i}}a-Guerrero, A. Zamudio-Ochoa, Y. Camacho-Villasana, R.
  Garc{\'{i}}a-Villegas, X. P{\'{e}}rez-Mart{\'{i}}nez, and A. Reyes-Prieto.
\newblock {Evolution of translation in mitochondria}.
\newblock In {\em Evolution of the Protein Synthesis Machinery and Its
  Regulation}, pages 109--142. Springer International Publishing, Cham, 2016.

\bibitem{Akabane2014}
S. Akabane, T. Ueda, K.~H. Nierhaus, and N. Takeuchi.
\newblock {Ribosome Rescue and Translation Termination at Non-standard Stop
  Codons by ICT1 in Mammalian Mitochondria}.
\newblock {\em {PLoS} Gen.}, 10(9):e1004616, 2014.

\bibitem{Kazemi2016}
M. Kazemi, F. Himo, and J. {\AA}qvist.
\newblock {Peptide Release on the Ribosome Involves Substrate-Assisted Base
  Catalysis}.
\newblock {\em ACS Catal.}, 6(12):8432--8439, 2016.\\
$\bullet$ Density functional theory probes several reaction mechanisms of peptide release and finds the one that is consistent with multiple experimental evidences.

\end{thebibliography}
\end{document}